\documentclass[twocolumn,showpacs,amsmath,amssymb,
nofootinbib,eqsecnum,floatfix]{revtex4-1}

\usepackage{graphicx}

\newlength{\dummysp}
\newcommand{\beq}{\begin{eqnarray}}
\newcommand{\eeq}{\end{eqnarray}}

\newcommand{\ord}[1]{{{\cal O}(#1)}}
\newcommand{\gappeq}{\mathrel{\rlap {\raise.5ex\hbox{$>$}}
{\lower.5ex\hbox{$\sim$}}}}
\newcommand{\lappeq}{\mathrel{\rlap{\raise.5ex\hbox{$<$}}
{\lower.5ex\hbox{$\sim$}}}}
\newcommand{\myref}[1]{(\ref{#1})}

\newcommand{\ben}{\begin{enumerate}}
\newcommand{\een}{\end{enumerate}}
\newcommand{\bit}{\begin{itemize}}
\newcommand{\eit}{\end{itemize}}

\def\[{\left [}
\def\]{\right ]}
\def\({\left (}
\def\){\right )}

\begin{document}

\title{Finite size scaling in minimal walking technicolor}

\author{Joel Giedt}
\email{giedtj@rpi.edu}
\author{Evan Weinberg}
\email{weinbe2@rpi.edu}
\affiliation{Department of Physics, Applied Physics and Astronomy,
Rensselaer Polytechnic Institute, 110 8th Street, Troy NY 12065 USA}

\date{Jan.~30, 2011}

\begin{abstract}
We compare observables to the finite
size scaling hypothesis in SU(2) lattice gauge theory
with two Dirac fermions in the adjoint representation.
The fits that we obtain yield an estimate of
the anomalous mass dimension that is consistent
with four loop perturbation theory: $\gamma = 0.51 \pm 0.16$,
with the error due to systematic uncertainties
in the finite size scaling analysis.  The result is
somewhat larger than one Schr\"odinger functional
study (by 1.3$\sigma$) but consistent with another.
\end{abstract}



\maketitle

\section{Introduction}

In technicolor models, the Higgs mechanism occurs through
condensation of new fermions that are subject to a gauge interaction
that is strong at the TeV scale \cite{Susskind:1978ms,Weinberg:1979bn}.  
Walking technicolor is a version of this theory that can 
suppress flavor-changing neutral currents by raising the
extended technicolor scale, while still having phenomenologically
acceptable Standard Model fermion masses, due to condensate
enhancement \cite{Holdom:1981rm,
Holdom:1984sk,Yamawaki:1985zg,Bando:1986bg,Appelquist:1986an,
Appelquist:1986tr,Appelquist:1987fc}.  
Higher representations
of the gauge group are believed to avoid problems with the S-parameter,
i.e.~electroweak precision constraints \cite{Eichten:1979ah,Lane:1989ej}.  
All of this has motivated the study of Minimal Walking 
Technicolor (MWTC) \cite{Sannino:2004qp}, which is SU(2) gauge theory
with two Dirac fermions in the adjoint (triplet) representation.

In order to study technicolor proposals nonperturbatively and
from first principles, several groups have been
using the techniques of lattice gauge theory; see the
review \cite{DelDebbio:2010zz} and references therein.  One of
the key questions is whether the theory ``walks'' (very slow
running of the coupling) or is attracted to an infrared
fixed point (IRFP).  An important quantity that can be
computed in the process of answering this question is
the anomalous mass dimension $\gamma$, which needs to
satify $\gamma \approx 1$ in order for the standard
walking technicolor picture to succeed.  (Alternatives
such as ``ideal walking'' are now being investigated
as improvements over the standard picture \cite{Fukano:2010yv}.)  
One of the ways
in which the lattice community has computed $\gamma$
is through the Schr\"odinger functional
method.  It was employed for SU(3) gauge group with sextet fermions
in \cite{Shamir:2008pb} and for MWTC in \cite{Bursa:2009we,DeGrand:2011qd}.
Analysis of the distribution of eigenvalues of the Dirac
operator has also been used \cite{Fodor:2008hm,DeGrand:2009hu,DelDebbio:2010ze}.  

An alternative approach is to compare observables computed
in lattice gauge theory (e.g., meson masses, the ``pion'' decay
constant) to the finite size scaling (FSS) hypothesis.  If the
theory is indeed driven to an IRFP, then the data on observables
should fit the FSS hypothesis.  Previous studies of FSS in 
lattice technicolor include \cite{DeGrand:2009hu,DelDebbio:2010hu,DelDebbio:2010hx,DeGrand:2011cu}.
Fits to the conformal hypothesis that assume a specific form
of the FSS function include \cite{Fodor:2011tu,Appelquist:2011dp},
where the infinite volume hyperscaling relation is imposed.
More general forms of the FSS function have also been considered
recently by the authors of \cite{Fodor:2011tu}, with the
result that for these forms the resulting conformal hypothesis
for SU(3) gauge group and 12 fundamental flavors has a low degree of
confidence in fitting the data \cite{KutiTalk}.
By contrast, \cite{DeGrand:2011cu} advocates an approach that does not impose
a specific form on the FSS function; this is one of the FSS methods
used in the earlier work \cite{DeGrand:2009hu}.  In this letter,
we apply this method to the case of MWTC in order to extract
an estimate of $\gamma$ under the assumption that an IRFP exists.

\section{Finite size scaling}
In the scaling regime, the correlation length will have
an asymptotic behavior dependent on the fermion mass $m$
with exponent $y_m$:
\beq
\xi \sim m^{-1/y_m}
\eeq
This exponent is related to the anomalous mass dimension
evaluated at the IRFP:
\beq
y_m = 1 + \gamma(g_*)
\eeq
It is a general consequence of the renormalization group equations
that the correlation length $\xi_L$ at finite size $L$ is given
by a scaling function of the infinite volume correlation length $\xi$
relative to $L$:
\beq
\xi_L/L = F(\xi/L)
\eeq
Thus we obtain the FSS formula in terms of fermion mass:
\beq
\xi_L/L = f(m L^{y_m})
\eeq
Corrections to scaling will be an important consideration for
us.  This translates into a correction that is appreciable for small $L$,
with an exponent $\omega$:
\beq
\xi_L/L = f(m L^{y_m}) + L^{-\omega} g(m L^{y_m})
\label{scavio}
\eeq
This form has also been considered in \cite{KutiTalk}; there it
was pointed out that fitting data to such a hypothesis would
require an extensive and highly accurate study.  For us the
main use of this equation is just that the scaling violations are
largest for the smallest values of $L$.  We use this as
an interpretation of data on small lattices that does not
fall on a scaling curve.  Our present study is not extensive
enough to fit to this more general form and extract $\omega$.
Below, we will consider $\xi_L = 1/M$ or $1/f_\pi$, where $M$ is a
meson mass.

\section{Fitting method}
The method described here seeks to optimize $y_m$ such
that all the data falls on a scaling curve.  It is
due to \cite{Bhatta} and was used in \cite{DeGrand:2009hu,DeGrand:2011cu}.
For each $L$ we have a data set $p$.  We use this
to obtain a fit $f_p$.  The types of fit functions
that we consider will be described below.  We then
use this fit function on the other values of $L$, which
we label as $L_j$. 

\begin{widetext}
We minimize the following function with respect to $y_m$.
\beq
P(y_m) = \frac{1}{N_{\text{over}}} \sum_p \sum_{j \not= p} \sum_{i,\text{over}}
\( \frac{\xi_L(m_{i,j})}{L_j} - f_p(L_j^{y_m} m_{i,j}) \)^2
\label{tomin}
\eeq
Here $i$ labels the different partially conserved axial current (PCAC) mass values for a given $L_j$.
The effect of this is to find a $y_m$ such that $f_p$ for the
other values $L_j, m_{i,j}$ is as close as possible to the
curve obtained from fitting $L_p, m_{i,p}$.  This is summed
over all possibilities $p$.  Also, ``over'' indicates that
only $i$ are used such that $m_{i,j} L_j^{y_m}$ falls within
the range of values of $m_{i,p} L_p^{y_m}$, so that the comparison
is to an interpolation of the $m_{i,p} L_p^{y_m}$ data,
rather than an extrapolation.  Unweighted fits were
used so that the approximation to the scaling curve
would pass through data at small $x=m L^{y_m}$,
where absolute (statistical) errors are largest.  (Using
a weighted fit reduces our conclusion for $\gamma$ by 4\%.)
\end{widetext}

\begin{table}
\begin{tabular}{|c|c|} \hline\hline
Type & $f(x)$ \\ \hline \hline
Quadratic & $c_0+c_1 x+c_2 x^2$ \\ \hline
Log quadratic & $c_0 + c_1 \ln x + c_2 (\ln x)^2$ \\ \hline
Piece-wise log-linear & Straight lines connecting data \\ \hline\hline
\end{tabular}
\caption{Interpolating functions that we use to fit data for
a fixed $L_p$.  In the last case, the straight lines
interpolating between data are on a semi-log plot. \label{tabfit}
}
\end{table}

For the fitting function we have considered the possibilities
listed in Table \ref{tabfit}.  In the case of the quadratic 
we follow one of the methods of \cite{DeGrand:2009hu,DeGrand:2011cu}.
The log quadratic fit was motivated by the behavior of
the data when $\xi_L/L$ is plotted versus $\ln(m L^{y_m})$,
which is close to a parabola.  The piece-wise log-linear
form was used as a third choice that trivially passes
through the data, giving a reasonable interpolation.

\section{Results}
We have used four observables:  the ``pion'' mass $m_\pi$,
the ``rho'' mass $m_\rho$, the ``$a_1$'' mass $m_{a_1}$,
and the ``pion'' decay constant $f_\pi$.  These are all
obtained from standard correlation functions using
point sources and sinks.  We fit the correlation functions
with a single exponential, allowing the first time $t_{\text{first}}$ in the fit
to be large enough for the excited state contributions
to be negligible.  This is determined by looking at the mass
of the meson as a function of $t_{\text{first}}$ and
extracting the value on the plateau.  Five values of
bare masses $m_0 a = -1.0, -1.1, -1.165, -1.175, -1.18$
on lattices of size $L/a = 10, 12, 16, 20, 24$ were
simulated, all at $\beta=2.25$.  These are the same configurations as were
generated in \cite{Giedt:2011kz}, and the values of
the PCAC mass and details on the simulations are given there.  
Also note that the size of the temporal direction is $T=2L$.

\begin{figure}
\includegraphics[height=2in,width=2in]{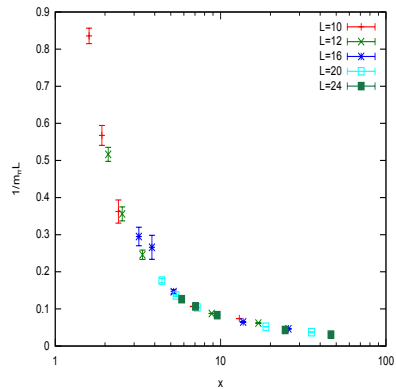}
\caption{Collapse of $\pi$ data for $y_m=1.46$.  Here and
in the other figures, $x=m L^{y_m}$.
\label{fig1} }
\end{figure}

\begin{figure}
\includegraphics[height=2in,width=2in]{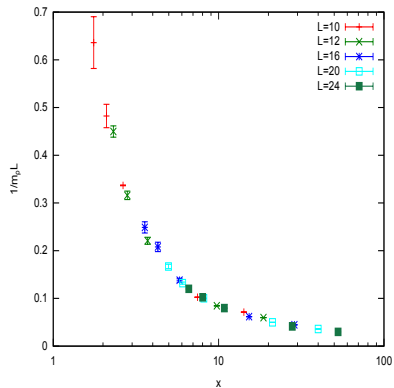}
\caption{Collapse of $\rho$ data for $y_m=1.50$.
\label{fig2} }
\end{figure}

\begin{figure}
\includegraphics[height=2in,width=2in]{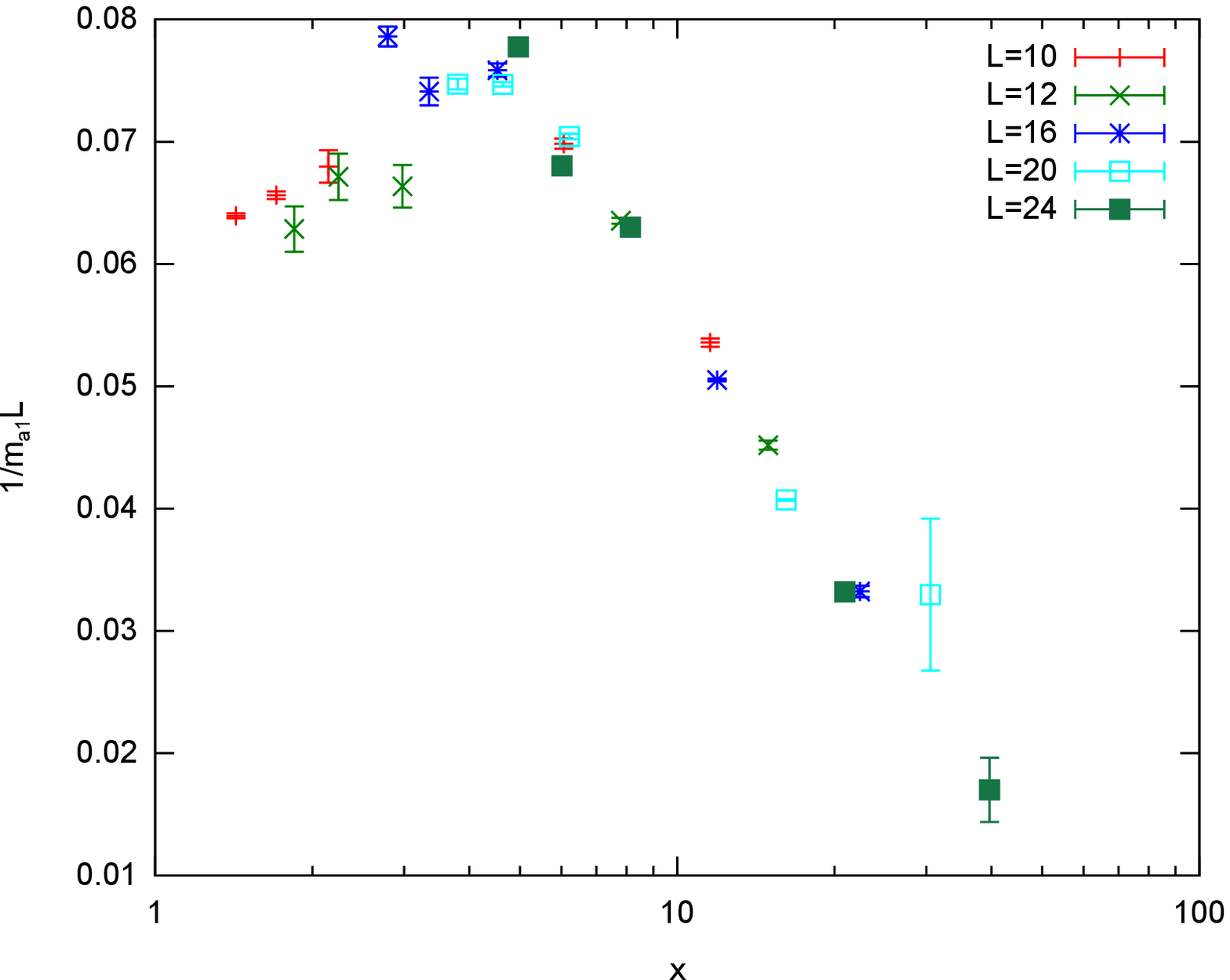}
\caption{The $a_1$ data for $y_m=1.41$.
The $L=10,12$ data do not fall on the curve, which
we interpret as large scaling violations.  
These points were excluded from the fit.
\label{fig3} }
\end{figure}

\begin{figure}
\includegraphics[height=2in,width=2in]{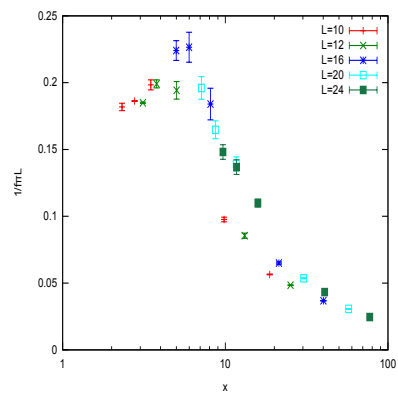} 
\caption{The $f_\pi$ data for $y_m=1.62$.
The $L=10,12$ data do not fall on the curve, which we
interpret as large scaling violations.  These points were excluded
from the fit.
\label{fig4} }
\end{figure}

Using these results, and performing the minimization
described in the previous section, we obtain values for $y_m$.
In the case of $m_{a_1}$ and $f_\pi$, the quantity
$\xi_L/L$ is small, and scaling violations [cf.~Eq.~\myref{scavio}] can compete
with the scaling function for small lattices.  For this
reason we exclude the small lattices $L/a=10,12$ for these
channels.  The results for $y_m$ are summarized in
Table \ref{tabus}.  It can be seen that each of the
channels, and each of the fitting methods are consistent
with each other within errors.  
The approximate collapse of data is shown in Figs.~\ref{fig1}-\ref{fig4};
it can be seen in Figs.~\ref{fig3} and \ref{fig4} that the excluded small lattice data does not fall on
the scaling curve.  We interpret this as being due to scaling violations,
though a thorough study extracting $\omega$ would be required to demonstrate this.
Another interpretation is that the theory does not have an IRFP,
and so the FSS fails for some channels.  It is also possible that
we are seeing the effect of $\beta=2.25$ not being close
enough to the fixed point coupling.  However we view the
collapse seen in Figs.~\ref{fig1} and \ref{fig2}
as favoring our scaling violation interpretation.

\begin{table}
\begin{center}
\begin{tabular}{|c|c|c|c|c|} \hline \hline
Observable & Quadratic	& Log Quad  &		PWL     &	Combined \\ \hline \hline
$m_\pi$	  &	1.67(93) & 1.26(54)	&  1.51(33) &	1.46(27) \\ \hline
$m_\rho$	& 1.67(88) & 1.37(39) &	 1.56(31) & 1.50(23) \\ \hline
$m_{a_1}$	&	1.40(52) & 1.42(27) &  1.41(22) &	1.41(16) \\ \hline
$f_\pi$   &	1.65(22) & 1.49(54) &	 1.60(29) &	1.62(17) \\ \hline \hline
\end{tabular}
\caption{The scaling exponent $y_m=1+\gamma$ for the various observables and methods of
interpolation.  In parentheses, jackknife error is shown, obtained from eliminating
one $m_{i,j}$ in all possible ways, in the minimization of \myref{tomin}.  
Because we use a large number of configurations, $\ord{10^3}$,
statistical error is negligible by comparison.
\label{tabus} }
\end{center}
\end{table}

We average the twelve values of $y_m$ for the four channels and
three fitting methods, weighted by the jackknife errors, to obtain
$\overline{\gamma}=0.51$.  The combined jackknife error from
the twelve fits is 0.097, and the standard deviation of
the twelve fits is 0.128.  We regard these as two sources
of systematic error and combine them in quadrature to
obtain a final estimate of
\beq
\gamma = 0.51 \pm 0.16
\eeq
In Table \ref{taball}, we compare this estimate with 
results from other groups using a variety of methods.
Our result is consistent with one of those obtained using
the Schr\"odinger functional \cite{Bursa:2009we}, perturbation theory, Monte Carlo
renormalization group
and the all-orders hypothesis of \cite{Pica:2010mt}.
We obtain a value significantly larger than the
one obtained in the FSS studies \cite{DelDebbio:2010hu,DelDebbio:2010hx} (1.8$\sigma$
difference) and somewhat larger than the Schr\"odinger 
functional study \cite{DeGrand:2011qd} (1.3$\sigma$
difference).

\begin{table}
\begin{tabular}{|c|c|} \hline \hline
Method & $\gamma$ \\ \hline \hline
SF \cite{Bursa:2009we} & $0.05 < \gamma < 0.56$ \\ \hline
SF \cite{DeGrand:2011qd} & $0.31 \pm 0.06$ \\ \hline
Perturbative 4-loop \cite{Pica:2010xq} & $0.500$ \\ \hline
Schwinger-Dyson \cite{Ryttov:2010iz} & $0.653$ \\ \hline
All-orders hypothesis \cite{Pica:2010mt} & $0.46$ \\ \hline
MCRG \cite{Catterall:2011zf} & $-0.6 < \gamma < 0.6$ \\ \hline
FSS \cite{DelDebbio:2010hu} & $0.05 < \gamma < 0.20$ \\ \hline
FSS \cite{DelDebbio:2010hx} & $0.22 \pm 0.06$ \\ \hline
FSS (here) & $0.51 \pm 0.16$ \\ \hline \hline
\end{tabular}
\caption{Summary of all MWTC results for the anomalous mass dimension.
SF is Schr\"odinger functional and MCRG is Monte Carlo renormalization group.
The perturbative result $\gamma=0.500$ was also given in the later, corrected version
of \cite{Ryttov:2010iz}, and relied on invariants calculated in \cite{Mojaza:2010cm}.
\label{taball} }
\end{table}

\section{Conclusions}
We have applied the FSS approach of \cite{DeGrand:2011cu} (one of
the approaches in \cite{DeGrand:2009hu}) to MWTC
and find values of the critical exponent that
are in agreement with perturbative results, but somewhat higher than
the Schr\"odinger functional result with the smallest
estimate of error \cite{DeGrand:2011qd}, by 1.3$\sigma$.
While there are significant systematic uncertainties, which we
interpret as being due to scaling violations on small volumes, the value of $\gamma$ is
too small for phenomenological models of condensate enhancement,
which requires $\gamma \approx 1$.
The complimentary information obtained by the present method suggests
that it be applied in other gauge theories of interest for conformal
or near-conformal dynamics.  Indeed we expect it to work in any
case for which the gauge coupling runs very slowly, so that fixed
point behavior is well approximated on the scales probed by the
study that is performed.  Unfortunately, as explained in \cite{DeGrand:2011cu},
a reasonable fit to the FSS assumption does not rule in or out
the existence of an IRFP, since all that is required
is a very slow running.  

We have highlighted some of the systematic uncertainties of
the method, and have illustrated how working on small volumes
hampers the effort to obtain an accurate $y_m$.  Future work includes simulations on
larger volumes so that $y_m$ can be obtained with greater
certainty.  Also, an improved lattice action should reduce
the size of the scaling violations, and we are currently
working in that direction for MWTC and other theories.

\section*{Acknowledgements}
This research was supported by the Dept.~of Energy,
Office of Science, Office of High Energy Physics, 
Grant No.~DE-FG02-08ER41575.  
We gratefully acknowledge the sustained use of
RPI computing resources over the course of a year,
both the on-campus SUR IBM BlueGene/L rack, as well
as continuous access to 1-4 racks of the sixteen
IBM BlueGene/L's situated at the Computational Center
for Nanotechnology Innovation.  We thank Tom DeGrand for
extensive discussions and comments.  We also thank
Luigi Del Debbio, Julius Kuti and Francesco Sannino 
for helpful comments.

\bibliography{fss}
\bibliographystyle{apsrev4-1}
\end{document}